\documentclass[prb,twocolumn,showpacs,amsmath,amssymb,superscriptaddress]{revtex4-1}

\usepackage{graphicx, amsmath, bm, braket, txfonts, color}

\begin{document}

\title{Strong spin-orbit interaction and helical hole states in Ge/Si nanowires}

\author{Christoph Kloeffel}
\affiliation{Department of Physics, University of Basel,
             Klingelbergstrasse 82, CH-4056 Basel, Switzerland}
\author{Mircea Trif}
\affiliation{Department of Physics, University of Basel,
             Klingelbergstrasse 82, CH-4056 Basel, Switzerland}
\affiliation{Department of Physics and Astronomy, University of California, Los Angeles, California 90095, USA}
\author{Daniel Loss}
\affiliation{Department of Physics, University of Basel,
             Klingelbergstrasse 82, CH-4056 Basel, Switzerland}


\begin{abstract}
We study theoretically the low-energy hole states of Ge/Si core/shell nanowires. The low-energy valence band is quasidegenerate, formed by two doublets of different orbital angular momenta, and can be controlled via the relative shell thickness and via external fields. We find that direct (dipolar) coupling to a moderate electric field leads to an unusually large spin-orbit interaction of Rashba type on the order of meV which gives rise to pronounced helical states enabling electrical spin control. The system allows for quantum dots and spin qubits with energy levels that can vary from nearly zero to several meV, depending on the relative shell thickness.  
\end{abstract}

\pacs{73.22.Dj, 72.25.Dc, 73.21.Hb, 73.21.La}

\maketitle

\section{Introduction}

Semiconducting nanowires are subject to intense experimental effort as promising candidates for single-photon sources,\cite{reimer:jnp11} field-effect transistors,\cite{xiang:nat06} and programmable circuits.\cite{yan:nat11} Progress is being made with both group-IV materials \cite{xiang:nat06, yan:nat11, lu:pnas05, hu:nna07} and III-V compounds, particularly InAs, where single-electron quantum dots \cite{fasth:prl07, schroer:arX11} (QDs) and universal spin-qubit control \cite{nadjperge:nat10} have been implemented. Proximity-induced superconductivity was demonstrated in these systems,\cite{doh:sci05, xiang:nna06} forming a platform for Majorana fermions.\cite{lutchyn:prl10, sau:prb10, oreg:prl10, alicea:nph11, mao:arX11, gangadharaiah:prl11}

The nanowires are operated in both the electron \cite{doh:sci05, fasth:prl07, schroer:arX11, nadjperge:nat10} (conduction band, CB) and hole \cite{hu:nna07, lu:pnas05, yan:nat11, xiang:nat06, xiang:nna06, quay:nph10} (valence band, VB) regimes. While these regimes are similar in the charge sector, holes can have many advantages in the spin sector. Due to strong spin-orbit interaction (SOI) 
on an atomic level, the electron spin is replaced by an effective spin $J = 3/2$, and even in systems that are inversion symmetric, the spin and momentum are strongly coupled, enabling efficient hole spin manipulation by purely electrical means. Holes, moreover, are very sensitive to confinement, which strongly prolongs their spin lifetimes.\cite{bulaev:prl05, heiss:prb07, trif:prl09, fischer:prb08, fischer:prl10, brunner:sci09} Also, VBs possess only one valley at the $\Gamma$ point, in contrast to the CBs of Ge and Si, which is particularly useful for spintronics devices such as spin filters \cite{streda:prl03} and spin qubits.\cite{loss:pra98} Most recently, spin-selective hole tunneling in SiGe nanocrystals was achieved.\cite{katsaros:arX11} 

In this paper, we analyze the hole spectrum of Ge/Si core/shell nanowires, which combine several useful features. The holes are subject to strong confinement in two dimensions and can be confined down to zero dimension (0D) in QDs.\cite{hu:nna07, roddaro:arX07, roddaro:prl08} Ge and Si can be grown nuclear-spin-free, and mean free paths around $0.5\mbox{ $\mu$m}$ have been reported.\cite{lu:pnas05} During growth, the core diameter ($\sim$5-100 nm) and shell thickness ($\sim$1-10 nm) can be controlled individually. The VB offset at the interface is large, $\sim$0.5 eV, so that holes accumulate naturally in the core.\cite{lu:pnas05, park:nlt10} Lack of dopants underpins the high mobilities \cite{xiang:nat06} and the charge coherence seen in 
proximity-induced superconductivity.\cite{xiang:nna06} 

We find that the low-energy spectrum in Ge/Si nanowires is quasidegenerate, in contrast to typical CBs. Static strain, adjustable via the relative shell thickness, allows lifting of this quasidegeneracy, providing a high degree of control. We also calculate the spectrum in longitudinal QDs, where this feature remains pronounced, which is essential for spin-qubit implementation. The nanowires are sensitive to external magnetic fields, with $g$ factors that depend on both the field orientation and the hole momentum. In particular, we find an additional SOI of Rashba type (referred to as direct Rashba SOI, \mbox{DRSOI}), which results from a direct dipolar coupling to an external electric field. This term arises in first order of the multiband perturbation theory, and thus is 10-100 times larger than the known Rashba SOI (RSOI) for holes which is a third-order effect.\cite{winkler:book} Moreover, the DRSOI scales linearly in the core diameter $R$ (while the RSOI is proportional to $R^{-1}$), so that spin-orbit interaction remains strong even in large nanowires. Similarly to the conventional Rashba SOI,\cite{quay:nph10, sau:prb10, oreg:prl10, lutchyn:prl10, alicea:nph11, mao:arX11, gangadharaiah:prl11, streda:prl03, braunecker:prb10, klinovaja:prl11} the DRSOI induces helical ground states, but with much larger spin-orbit energies (meV range) than in other known semiconductors.

The paper is organized as follows. In Sec.\ \ref{sec:ModelHamiltonian} we introduce the unperturbed Hamiltonian for holes inside the Ge core and provide its exact, numerical solution. The system is very well described by an effective 1D Hamiltonian, which we derive in Sec.\ \ref{sec:Eff1DHamiltonian}. In Sec.\ \ref{sec:StaticStrain} we include the static strain and find a strong dependence of the nanowire spectrum on the relative shell thickness. The spectrum of Ge/Si-nanowire-based QDs is discussed subsequently (Sec.\ \ref{sec:QDspectrum}). In the main section, \mbox{Sec.\ \ref{sec:DRSOI}}, we analyze the hole coupling to electric fields and compare the DRSOI to the standard RSOI. In this context, we also show that Ge/Si nanowires present an outstanding platform for helical hole states and Majorana fermions. Magnetic field effects are discussed in Sec.\ \ref{sec:MagneticField}, followed by our summary and final remarks, Sec.\ \ref{sec:Discussion}. Technical details and additional information are appended. 

\begin{figure}[tb]
\begin{center}
\includegraphics[width=0.78\linewidth]{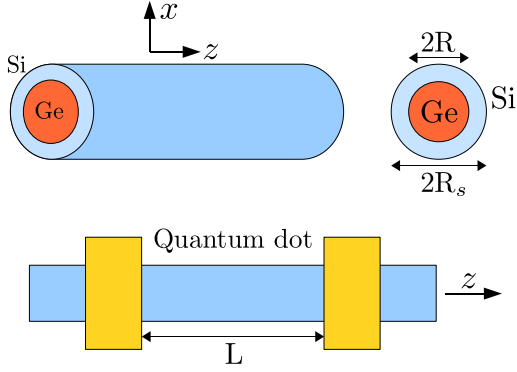}
\caption{Schematic drawing of the systems studied in this paper. Top: Excerpt of a Ge/Si nanowire with core radius $R$ and shell thickness $R_s - R$, where the $z$ axis corresponds to the axis along the wire. The nanowires are typically several micrometers in length and can therefore be considered infinitely extended, hosting a 1D hole gas inside their cores. The surrounding Si shell influences the hole spectrum through static strain. Bottom: Quantum dots of (effective) length $L$ form when the holes are subject to additional confinement in the $z$ direction. This can be realized via gates (Refs.\ \onlinecite{hu:nna07, roddaro:arX07, roddaro:prl08}) or, in principle, by surrounding the Ge with layers of barrier material during growth (Ref.\ \onlinecite{wen:sci09}).}
\label{schemeWireQD}
\end{center}
\end{figure}

\section{Model Hamiltonian and numerical solution}
\label{sec:ModelHamiltonian}

In cubic semiconductors, the VB states are well described by the  Luttinger-Kohn (LK) Hamiltonian,\cite{luttinger:pr56, commentLK}
\begin{equation}
H_{\rm LK} = \frac{\hbar^2}{2m} \left[ \left(\gamma_1+\frac{5}{2}\gamma_s\right)k^2 - 2\gamma_s(\bm{k}\cdot\bm{J})^2 \right],
\label{LK_spherical}
\end{equation} 
where $J_{x,y,z}$ (in units of $\hbar$) are the three components of the effective electron spin in the VB, $m$ is the bare electron mass, $\hbar \bm{k}$ is the momentum operator, and $\gamma_1$ and $\gamma_s \equiv (2\gamma_2+3\gamma_3)/5$ are the Luttinger parameters in spherical approximation, which is well applicable for Ge ($\gamma_1 = 13.35$, $\gamma_s = 5.11$).\cite{lawaetz:prb71} In studying nanowires [Fig.\ \ref{schemeWireQD} (top)], the LK Hamiltonian must be supplemented with the confinement in the transverse directions ($x$-$y$ plane), perpendicular to the wire axis $z$. Since we are interested in the low-energy states, we can add two more simplifications at this stage. First, since the low-energy states are located near the core center, we can assume a potential with cylindrical symmetry even though the real system is not perfectly symmetric. Second, due to the large VB offset, the confinement can be treated as a hard wall,
\begin{equation}
V(r) = \left\{ \begin{array}{lr}
0, & r < R , \\
\infty, & r > R ,
\end{array} \right.
\end{equation}
with $R$ as the core radius. Given this confinement, the total Hamiltonian $H_{\rm LK} + V$ commutes with the operator $F_z = L_z + J_z$, where $L_z = -i \partial_\phi$ is the orbital angular momentum along the wire axis, so that $F_z$ is a good quantum number and the states can be classified accordingly.\cite{sercel:prb90, csontos:prb09} The system is also time-reversal symmetric (Kramers doublets), and due to cylindrical symmetry one obtains the same spectrum for the same $|F_z|$. This is valid for any circular confinement and does not require the assumption of a hard wall. We note that, again in clear contrast to the CB case, $L_z$ is not conserved in the VB. 

\begin{figure}[tb]
\begin{center}
\includegraphics[width=0.82\linewidth]{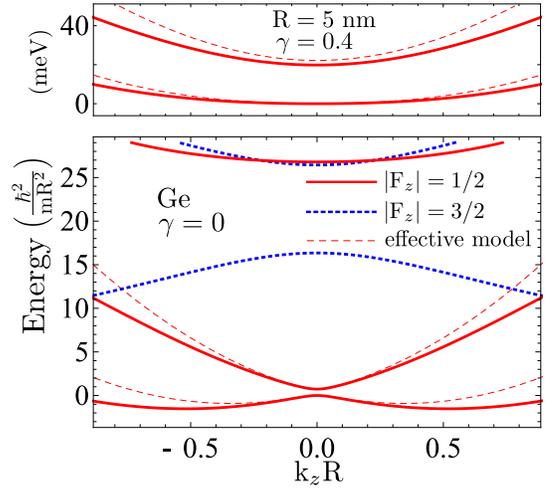}
\caption{Low-energy hole spectrum of a Ge nanowire as a function of the longitudinal wave number $k_z$. In the unstrained case, $\gamma = 0$, the plot is independent of $R$, with $\hbar^2/(m R^2)\simeq 0.76\mbox{ meV}$ for $R = 10\mbox{ nm}$. Due to time-reversal invariance and cylindrical symmetry, each line is a twofold degeneracy, where red (blue) indicates quantum numbers $F_z = \pm 1/2$ ($F_z = \pm 3/2$). At $k_z = 0$ the spectrum is quasidegenerate, with the lowest states having $L_z\approx 0$ (ground states) and $|L_z| = 1$ (excited states) character. Dashed red lines result from the effective 1D model for the lowest subspace, where $k_z$ is treated perturbatively. The top figure is a plot of the low-energy sector of a strained system, $\gamma = 40\%$, illustrating strong dependence on the Si shell thickness.}
\label{germaniumSpectrum}
\end{center}
\end{figure} 

The Hamiltonian separates into $4\times 4$ blocks corresponding to given $F_z$. By solving $H_{\rm LK} + V$ numerically, using an ansatz analogous to those in Refs.\ \onlinecite{sercel:prb90, csontos:prb09}, we find that the low-energy spectrum in the Ge core is formed by two quasidegenerate bands, with $F_z = \pm 1/2$ each, where the ground (excited) states at $k_z = 0$ are of $L_z \approx 0$ ($|L_z|=1$) type. These four (in total) bands are well separated from higher bands, and the quasidegeneracy indicates that one can project the problem onto this subspace. A plot of the spectrum is shown in Fig.\ \ref{germaniumSpectrum} (bottom). 

\section{Effective 1D Hamiltonian}
\label{sec:Eff1DHamiltonian}
The present analysis does not, however, allow us to derive an effective 1D Hamiltonian describing the lowest-energy states. For this, we integrate out the transverse motion and treat $k_z$ in perturbation theory ($k_z R < 1$). The four eigenstates $g_{\mp}$ and $e_{\pm}$, corresponding to ground and excited states for $F_z = \pm 1/2$ at $k_z = 0$, serve as the basis states in the effective 1D Hamiltonian. The subscript refers to the sign of the contained spin state $\ket{\pm 3/2}$, since the system at $k_z = 0$ can be separated into two $2 \times 2$ spin blocks;\cite{csontos:prb09} details of the calculation are described in Appendix \ref{app:BasisStates1D}. Knowledge of $g_\pm$ and $e_\pm$, with eigenenergies $E_g \equiv 0$ and $E_e \equiv \Delta$, allows us to include the $k_z$-dependent terms of the LK Hamiltonian. The diagonal matrix elements take on the form $\bra{g_\pm}H_{\rm LK}\ket{g_\pm} = \hbar^2 k_z^2/(2m_{g})$, $\bra{e_\pm}H_{\rm LK}\ket{e_\pm} = \hbar^2 k_z^2/(2m_{e}) + \Delta$, and the nonzero off-diagonal terms are of type $\bra{e_\pm}H_{\rm LK}\ket{g_\mp} = i C k_z$, with $C$ as a real-valued coupling constant.\cite{commentPhases} Summarized in matrix notation, this yields
\begin{equation}
H_{\rm LK}^{\rm eff} = A_{+} + A_{-} \bm{\tau}_z + C k_z \bm{\tau}_y \bm{\sigma}_x ,
\label{eff_ham}
\end{equation}
where $A_{\pm} \equiv \hbar^2 k_z^2 (m_g^{-1} \pm m_e^{-1})/4 \pm \Delta/2$, and $\bm{\tau}_i$, $\bm{\sigma}_i$ are the Pauli matrices acting on $\{g, e\}$, $\{+, -\}$ (see also Appendix \ref{app:Representation}).
For Ge, the values are $\Delta = 0.73\mbox{ }\hbar^2/(mR^2)$, $C = 7.26\mbox{ }\hbar^2/(mR)$, $m_g \simeq m/(\gamma_1+2\gamma_s) = 0.043\mbox{ }m$, and $m_e=m/(\gamma_1+\gamma_s) = 0.054\mbox{ }m$. The eigenspectrum 
\begin{equation}
E_{g,e}(k_z) = A_+ \mp \sqrt{A_-^2 + C^2 k_z^2}
\end{equation}
nicely reproduces all the key features of the exact solution and is added to Fig.\ \ref{germaniumSpectrum} for comparison, with good agreement for $k_z R < 1$. 

\section{Static strain}
\label{sec:StaticStrain}
To the above model one needs to add the effects of static strain, since the Si shell (radius $R_s$) tends to compress the Ge lattice. A detailed derivation of the strain field in Ge/Si core/shell nanowires will be provided elsewhere; here we just quote the results needed to calculate the hole spectrum. Coupling is described by the Bir-Pikus Hamiltonian $H_{\rm BP}$, \mbox{Eq.\ (\ref{BirPikusOfficial})}, which for Ge (the spherical approximation applies) is of the same form as Eq.\ (\ref{LK_spherical}), with $k_i k_j$ replaced by the strain tensor elements $\epsilon_{ij}$.\cite{birpikus:book} Assuming a stress-free wire surface and continuous displacement and stress at the interface, symmetry considerations and Newton's second law require $\epsilon_{xx} = \epsilon_{yy}$ and $\epsilon_{xy} = \epsilon_{xz} = \epsilon_{yz} =0$ within the core, so that only terms proportional to $J_z^2$ contribute. Hence, $F_z$ remains a good quantum number, $[H_{\rm LK} + V + H_{\rm BP}, F_z] = 0$, which allows us to solve the system exactly even in the presence of strain, following the same steps as described in Sec.\ \ref{sec:ModelHamiltonian}. It is important that these exact spectra show that the low-energy states [Fig.\ \ref{germaniumSpectrum} (bottom)] separate even further from the higher bands when the Ge core is strained by a Si shell, so that the low-energy sector remains energetically well isolated and projection onto this subspace is always valid.  

In the 1D model, strain leads to a simple rescaling of the energy splitting $\Delta \to \Delta + \delta(\gamma)$, where $0 \leq \delta(\gamma) \lesssim 30\mbox{ meV}$ for $0\leq \gamma < \infty$, with $\gamma \equiv (R_s - R)/R$ as the relative shell thickness. Hence, $\delta$ is independent of the core radius, while $\Delta \propto R^{-2}$. We note that $\Delta \simeq 0.6 \mbox{ meV}$ for a wire of $R=10\mbox{ nm}$, which makes this energy scale very small. Therefore the splitting can be changed not only via $R$, but also via $R_s$. In fact, the system can be varied from the quasidegenerate to an electronlike regime [Fig.\ \ref{germaniumSpectrum} (top)], where the $L_z \simeq 0$ and $|L_z| \simeq 1$ states are parabolas. 

\begin{figure}[tb]
\begin{center}
\includegraphics[width=1.00\linewidth]{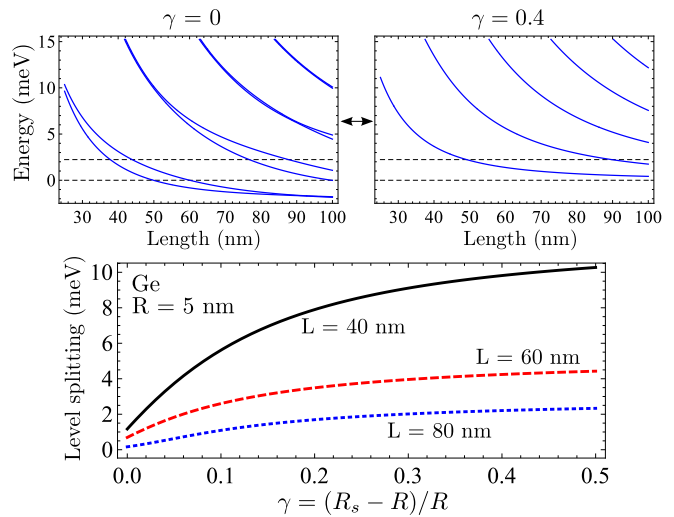}
\caption{Top: Hole energy spectrum in a nanowire-based QD (Ge/Si core/shell, $R = 5\mbox{ nm}$), for both a thin and a thick shell, as a function of confinement length $L$. Each line corresponds to a Kramers pair, and dashed lines represent $\Delta$ for comparison. Bottom: Level splitting of the two lowest Kramers doublets as a function of relative shell thickness $\gamma$ and for different lengths $L$. Static strain, induced via the shell, allows continuous tuning of the energy gap over several meV, an attractive feature for spin-qubit applications. For details, see Appendix \ref{app:QuantumDots}.}
\label{QDsplitting}
\end{center}
\end{figure}

\section{Quantum dot spectrum}
\label{sec:QDspectrum}
We analyze this feature in more detail by calculating the eigenenergies of Ge/Si-nanowire-based QDs [Fig.\ \ref{schemeWireQD} (bottom)]. All steps of this calculation are carefully explained in Appendix \ref{app:QuantumDots}. Remarkably, the variability with $R_s$ also transfers to the QD levels. Figure \ref{QDsplitting} shows the spectrum as a function of confinement length for a wire with both thin and thick shells and plots the energy splitting of the lowest Kramers doublets as a function of $\gamma$. For a negligible shell, the states lie so close in energy that additional degeneracies may even be observed. With increasing $R_s$, the QD spectrum changes monotonically from the quasidegenerate regime to gaps of several meV, which should, in particular, be useful for implementing spin qubits.

\section{Direct Rashba SOI and helical hole states}
\label{sec:DRSOI}
\begin{figure}[tb]
\begin{center}
\includegraphics[width=0.85\linewidth]{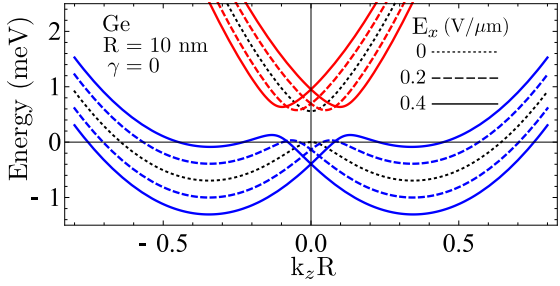}
\caption{Dispersion relation for holes in a Ge nanowire of $R = 10\mbox{ nm}$, negligible shell, and an applied electric field $E_x$ along $x$, 
calculated from $H_{\rm LK}^{\rm eff} + H_{\rm DR}$, Eqs.\ (\ref{eff_ham}) and (\ref{eff_ham_electric}), with $H_{\rm DR}$ as the DRSOI Hamiltonian. Hole bands of lower (higher) energy are plotted blue (red). The RSOI is about 100 times smaller than the DRSOI and thus negligible. Note that the DRSOI shows qualitatively similar features to the standard Rashba SOI with dispersion curves shifted along $k_z$ against each other.}
\label{plot_SOIx_Ex}
\end{center}
\end{figure}

An electric field $E_x$ applied along $x$ couples directly to the charge of the hole via the dipole term 
\begin{equation}
H_{\rm ed} = -e E_x x,
\end{equation}
with $x = r \cos(\phi)$ as the carrier position in field direction. For holes in the Ge core we expect this energy gradient to have sizable effects compared to electron systems, since the low-energy band is made of quasidegenerate states of different $L_z$ character. Moreover, $E_x$ will also couple directly to the spins due to the SOI in the VB. Projection of $H_{\rm ed}$ onto the subspace yields the effective SOI Hamiltonian
\begin{equation}
H_{\rm DR} = H_{\rm ed}^{\rm eff} = e E_x U \bm{\tau}_x \bm{\sigma}_z ,
\label{eff_ham_electric}
\end{equation}
referred to as direct Rashba SOI (DRSOI), characterized by the coupling constant $U = \bra{g_{+}}(-x)\ket{e_{+}}$. The form of \mbox{Eq.\ (\ref{eff_ham_electric})} still resembles that in the CB case, where dipolar coupling cannot modify the spins. However, the additional $k_z \bm{\tau}_y \bm{\sigma}_x$ term in $H_{\rm LK}^{\rm eff}$ makes the key difference to the CB and accounts for the SOI featured in the LK Hamiltonian. Indeed, by diagonalizing $H_{\rm LK}^{\rm eff} + H_{\rm DR}$ we find that the DRSOI lifts the twofold degeneracy, as plotted in Fig.\ \ref{plot_SOIx_Ex}. Surprisingly, the effects closely resemble a standard RSOI for holes in a transverse electric field (see discussion below). [Again, this is not the case for the CB, where $H_{\rm ed}$ does not lift the degeneracy since spin and orbit are decoupled (in leading order).]

\begin{figure}[tb]
\begin{center}
\includegraphics[width=0.85\linewidth]{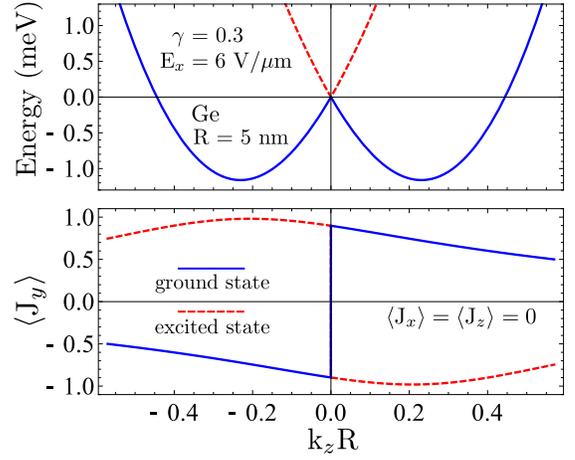}
\caption{Top: Splitting of the lowest valence band when 
an electric field $E_x = 6 \mbox{ V/$\mu$m}$ is applied to a Ge/Si nanowire of  $R = 5\mbox{ nm}$ and $R_s = 6.5\mbox{ nm}$. Ground (excited) hole states are plotted blue (red). $E_{\rm SO} > 1.0 \mbox{ meV}$, a large value compared to that for InAs (Refs.\ \onlinecite{fasth:prl07, dhara:09}), and the degeneracy at $k_z = 0$ may be lifted via a magnetic field (see Fig.\ \ref{HelicalStateWithB}). The conventional RSOI for holes is negligible.
Bottom: Plot of $\langle J_y \rangle$ for the above system, where $\langle J_x \rangle$ and $\langle J_z \rangle$ are zero throughout. In the ground state, the nanowire carries opposite spins in opposite directions with $|\langle J_y \rangle|\geq 1/2$. }
\label{HelicalStateNoB}
\end{center}
\end{figure}

As a consequence, when analyzing the eigenstates of $H_{\rm LK}^{\rm eff} + H_{\rm DR}$ for their spin properties, we find that an electric field generates helical ground states, i.e., holes of opposite spin move in opposite directions. Figure \ref{HelicalStateNoB} (top) shows the splitting of the lowest band when $E_x = 6 \mbox{ V/$\mu$m}$ is applied to a typical Ge/Si nanowire of 5 nm core radius and 1.5 nm shell thickness. Even though RSOI is absent, the result resembles the typical CB spectra considered in previous studies, where Rashba SOI for electrons leads to two horizontally shifted parabolas in the $E$-$k$ diagram.\cite{quay:nph10, sau:prb10, oreg:prl10, alicea:nph11, streda:prl03, braunecker:prb10} Moreover, the analogy also holds for the spins, which are twisted toward the $y$ direction, perpendicular to both the propagation axis $z$ and the field direction $x$. As Fig.\ \ref{HelicalStateNoB} (bottom) illustrates, $\langle J_y \rangle$ in the ground state is an antisymmetric function of $k_z$, the characteristic feature of a helical mode. We note that $\langle J_x \rangle = \langle J_z \rangle = 0$ throughout, so that the spins are indeed oppositely oriented. The values of $|J_y|$ around the band minima are $\geq$1/2, while the spin-orbit (SO) energy, i.e., the difference between band minimum and degeneracy at $k_z = 0$, is $E_{\rm SO} > 1.0$ meV. This value exceeds the reported $100 \mbox{ $\mu$eV}$ in InAs nanowires by a factor of 10 (see also Appendix \ref{app:SOenergyInAsNanowires}),\cite{fasth:prl07, dhara:09} and further optimization is definitely possible via both the gate voltage and the shell thickness.

We can understand the qualitative similarity of the DRSOI, Eq.\ (\ref{eff_ham_electric}), and RSOI,\cite{winkler:book} 
\begin{equation}
H_{\rm SO} = \alpha E_x(k_yJ_z-k_zJ_y) ,
\label{rashbaTermMainText}
\end{equation}
by projecting the latter onto the low-energy subspace, which yields 
\begin{equation}
H_{\rm R} = H_{\rm SO}^{\rm eff} \simeq \alpha E_x S \bm{\tau}_x \bm{\sigma}_z
\label{eff_ham_Rashba}
\end{equation}
for $k_z R < 1$, with $S = \bra{g_{+}}k_y J_z\ket{e_{+}}$. Further information on $H_{\rm SO}$, $H_{\rm R}$, and the Rashba coefficient $\alpha$ can be found in Appendix \ref{app:RSOIandCoefficient}. This formal analogy of $H_{\rm DR}$ and $H_{\rm R}$, Eqs.\ (\ref{eff_ham_electric}) and (\ref{eff_ham_Rashba}), immediately implies that Ge/Si nanowires provide a promising platform for novel quantum effects based on Rashba-type SOI.\cite{schroer:arX11, nadjperge:nat10, quay:nph10, sau:prb10, oreg:prl10, lutchyn:prl10, mao:arX11, gangadharaiah:prl11, alicea:nph11, streda:prl03, braunecker:prb10, klinovaja:prl11} A particular advantage of the DRSOI, as compared to conventional Rashba SOI, is its unusually large strength. While the  Rashba term for holes arises in third order of multiband perturbation theory and thus scales with $1/(\mbox{band gap})^{2}$, the DRSOI is a first-order effect and therefore much stronger.\cite{winkler:book} Explicit values for Ge are $U=0.15\mbox{ }R$, $S=0.36/R$, and $\alpha \approx -0.4 \mbox{ nm$^2 e$}$, so that, in typical nanowires with $R= \mbox{5-10 nm}$, $H_{\rm DR}$ dominates $H_{\rm R}$ by one to two orders of magnitude (Appendix \ref{app:RSOIandCoefficient}). Moreover, sizable RSOI would require unusually small confinement, since $H_{\rm R} \propto R^{-1}$. In stark contrast, for DRSOI we find $H_{\rm DR} \propto R$, which allows one to realize the desired coupling strengths in larger wires as well. The upscaling, however, is limited by the associated decrease of level splitting ($\propto R^{-2}$) and of the term $C k_z \bm{\tau}_y \bm{\sigma}_x$ ($ \propto R^{-1}$) in Eq.\ (\ref{eff_ham}).
  
\section{Magnetic field effects}
\label{sec:MagneticField}

\begin{figure}[tb]
\begin{center}
\includegraphics[width=0.85\linewidth]{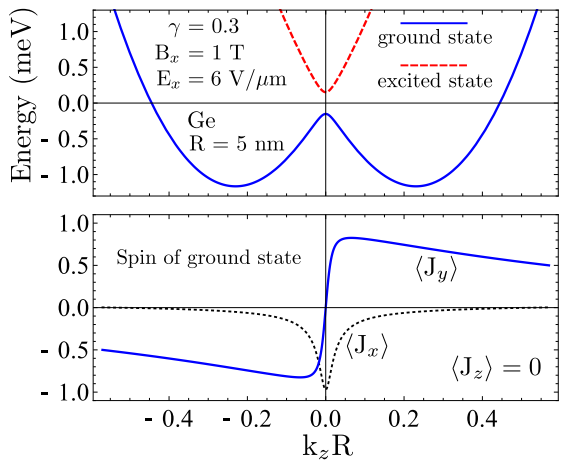}
\caption{Top: Hole spectrum of Fig.\ \ref{HelicalStateNoB} (top) in the presence of $B_x = 1 \mbox{ T}$. The magnetic field opens a gap of $0.30 \mbox{ meV}$ at $k_z = 0$, corresponding to a $g$ factor above 5.
Bottom: Plot of the ground state spin, $\langle J_x \rangle$ and $\langle J_y \rangle$, where $\langle J_z \rangle =0$ throughout. At energies within the gap, the Ge/Si nanowire features helical hole states with $E_{\rm SO} > 1.0 \mbox{ meV}$, $|k_z| \simeq 90\mbox{ $\mu$m}^{-1}$, and $|\langle J_y \rangle|\geq 1/2$.}
\label{HelicalStateWithB}
\end{center}
\end{figure}

The Kramers degeneracy can be lifted by an external magnetic field $\bm{B}$, which couples to the holes in two ways, first, via the orbital motion, through the substitution $\hbar\bm{k} \rightarrow -i\hbar\bm{\nabla} + e\bm{A}(\bm{r})$, with $\bm{A}(\bm{r})$ as the vector potential, and second, via the Zeeman coupling $H^Z_{\bm{B}} = 2\kappa\mu_B\bm{B}\cdot\bm{J}$, where $\kappa$ is a material parameter. For $\bm{B}$ along $z$ ($x$), parallel (perpendicular) to the wire, the 1D Hamiltonian is of the form
\begin{eqnarray}
H_{{\rm B},\hspace{0.02cm} z} &=& \mu_B B_z \left(Z_1 \bm{\sigma}_z + Z_2 \bm{\tau}_z \bm{\sigma}_z + Z_3 k_z \bm{\tau}_x \bm{\sigma}_y\right), \label{effBalongZ} \\
H_{{\rm B},\hspace{0.02cm} x} &=& \mu_B B_x \left(X_1 \bm{\sigma}_x + X_2 \bm{\tau}_z \bm{\sigma}_x + X_3 k_z \bm{\tau}_y\right), \label{effBalongX}
\end{eqnarray}
where the real-valued constants $Z_i$ ($X_i$) are listed in Eq.\ (\ref{bFieldZiAndXi}) of Appendix \ref{app:MagneticField}. The results agree with recent experiments, where the $g$ factors in Ge/Si-nanowire-based QDs (multihole regime) were found to vary dramatically with both the orientation of $\bm{B}$ and also the QD confinement.\cite{roddaro:arX07, roddaro:prl08} In the absence of electric fields, the ground state $g$ factor $g_\parallel(k_z)$ for $B_z$ along the wire turns out to be small for $k_z = 0$, $|g_\parallel(0)| \simeq 0.1$, and increases as $|k_z|$ increases. In contrast, the $g$ factor $g_\perp(k_z)$ for a perpendicular field $B_x$ is large at $k_z = 0$, $|g_\perp(0)| \simeq 6$, and decreases as $|k_z|$ increases, until $g_\perp(k_z)$ eventually changes sign at $|k_z| \approx 0.5/R$. We note that these results for the ground state cannot be directly compared to experimental results in the multihole regime, as the $g$ factors in the excited state already show a clearly different dependence on $k_z$. In the presence of an electric field $E_x$, the effective $g_\parallel$ and $g_\perp$ at $k_z = 0$ may, to some extent, be tuned by the strength of $E_x$.

Detailed analysis of the low-energy Hamiltonian yields the result that the combination of magnetic and electric fields allows for optimal tuning of the energy spectrum. For instance, $B_x = 1 \mbox{ T}$ opens a gap of $0.30\mbox{ meV}$ at $k_z = 0$ in Fig.\ \ref{HelicalStateNoB} (top), keeping the spin properties for $k_z \neq 0$ unaffected. This corresponds to $|g_\perp(0)| \simeq 5.2$ and is illustrated in Fig.\ \ref{HelicalStateWithB}. With the Fermi level within the induced gap, the spectrum of Fig.\ \ref{HelicalStateWithB} presents a promising basis for applications using helical hole states. Remarkably, an all-perpendicular setup with, e.g., $B_x$ along $x$ and $E_y$ along $y$, $H_{{\rm DR},\hspace{0.02cm} y} = - e E_y U \bm{\tau}_y$, leads to an asymmetric spectrum where only states with one particular direction of motion may be occupied, which moreover provide a well-polarized spin along the magnetic field axis. As before, this does not require standard RSOI.

\section{Discussion}
\label{sec:Discussion}
The low-energy properties found in this work make Ge/Si core/shell nanowires promising candidates for applications. The dipole-induced formation of helical modes proves useful for several reasons. First, the strength and orientation of externally applied electric fields are well controllable via gates. Second, the DRSOI scales linearly in $R$, instead of $R^{-1}$, and thicker wires remain operational. Third, the system is sensitive to magnetic fields, and undesired degeneracies at $k_z = 0$ may easily be lifted, with $|g_\perp(0)| \gtrsim 5$. Finally, helical modes with large $E_{\rm{SO}}$ and wave numbers $k_F$ are achievable using moderate electric fields of order V/$\mu$m. In Fig.\ \ref{HelicalStateWithB}, with the Fermi level inside the gap opened by the magnetic field, these are $E_{\rm SO} > 1.0 \mbox{ meV}$ and $k_F \simeq 90\mbox{ $\mu$m}^{-1}$, with $|\langle J_y \rangle|\geq 1/2$, and optimization via both the gate voltage and the Si shell is possible. For $R = 10\mbox{ nm}$ and thin shells, due to the quasidegeneracy at $\gamma \to 0$, even small electric fields of $\sim$0.1$\mbox{ V/$\mu$m}$ are sufficient to form helical states with $E_{\rm SO} \gtrsim 0.3\mbox{ meV}$. We note that a strong SOI, tuned via electric fields, was recently reported for Ge/Si nanowires based on magnetotransport measurements.\cite{hao:nlt10} 

The nanowire spectrum can be changed from the quasidegenerate to an electronlike regime, depending on the shell thickness. This moreover holds for QD spectra, so that, given the strong response to electric and magnetic fields, Ge/Si wires also seem attractive for applications in quantum information processing, particularly via electric-dipole-induced spin resonance.\cite{nadjperge:nat10, schroer:arX11, golovach:prb06} Finally, when combined with a superconductor,\cite{xiang:nna06} the DRSOI in these wires provides a useful platform for Majorana fermions.\cite{lutchyn:prl10, sau:prb10, oreg:prl10, mao:arX11, gangadharaiah:prl11, alicea:nph11} 

\acknowledgments{
We thank C.\ Marcus, Y.\ Hu, and F.\ Kuemmeth for helpful discussions and acknowledge support from the Swiss NF, NCCRs Nanoscience and QSIT, DARPA, and the NSF under Grant No.\ DMR-0840965 (M.T.).
}


\appendix

\section{Representation of spin matrices}
\label{app:Representation}
All results presented in this paper are based on the following representation of the spin-3/2 matrices:
\begin{gather}
J_x = \left( \begin{array}{cccc}
0 & \frac{\sqrt{3}}{2} & 0 & 0\\
\frac{\sqrt{3}}{2} & 0 & 1 & 0\\
0 & 1 & 0 & \frac{\sqrt{3}}{2}\\
0 & 0 & \frac{\sqrt{3}}{2} & 0 \end{array} \right), \\
J_y = \left( \begin{array}{cccc}
0 & -i\frac{\sqrt{3}}{2} & 0 & 0\\
i\frac{\sqrt{3}}{2} & 0 & -i & 0\\
0 & i & 0 & -i\frac{\sqrt{3}}{2}\\
0 & 0 & i\frac{\sqrt{3}}{2} & 0 \end{array} \right), \\
J_z = \left( \begin{array}{cccc}
\frac{3}{2} & 0 & 0 & 0\\
0 & \frac{1}{2} & 0 & 0\\
0 & 0 & -\frac{1}{2} & 0\\
0 & 0 & 0 & -\frac{3}{2} \end{array} \right).
\end{gather}
The Pauli operators $\bm{\tau}_i$ (referring to $\{g,e\}$) and $\bm{\sigma}_i$ (acting on $\{+,-\}$) are defined as
\begin{gather}
\bm{\tau}_x = \left( \begin{array}{cc}
0 & 1 \\
1 & 0
\end{array} \right), \mbox{ }\mbox{ }
\bm{\tau}_y = \left( \begin{array}{cc}
0 & -i \\
i & 0
\end{array} \right), \mbox{ }\mbox{ }
\bm{\tau}_z = \left( \begin{array}{cc}
1 & 0 \\
0 & -1
\end{array} \right),
\end{gather} 
and analogously for $\bm{\sigma}_i$.

\section{Basis states for the effective 1D Hamiltonian}
\label{app:BasisStates1D}
In this appendix we outline the calculation of the basis states $\{g_{+}, g_{-}, e_{+}, e_{-} \}$. For $k_z = 0$, each of the $4\times 4$ blocks for given quantum number $F_z$ and energy $E$ reduces to two $2\times 2$ blocks, labeled by $\pm$ according to the sign of the contained spin state $\ket{\pm 3/2}$. In the absence of confinement, using an ansatz analogous to those in Refs.\ \onlinecite{sercel:prb90, csontos:prb09}, the eigenstates to be considered are
\begin{eqnarray}
\psi^{F_z}_{{\rm hh},\pm} &=& J_{F_{z}\mp3/2}(k_{\rm hh}r)e^{i(F_z\mp 3/2)\phi}\ket{\pm 3/2} \nonumber \\ & & - \sqrt{3}J_{F_{z}\pm 1/2}(k_{\rm hh}r)e^{i(F_z\pm 1/2)\phi}\ket{\mp 1/2} ,\\
\psi^{F_z}_{{\rm lh},\pm} &=& \sqrt{3}J_{F_{z}\mp 3/2}(k_{\rm lh}r)e^{i(F_z\mp 3/2)\phi}\ket{\pm 3/2} \nonumber \\ & & + J_{F_{z}\pm 1/2}(k_{\rm lh}r)e^{i(F_z\pm1/2)\phi}\ket{\mp 1/2},
\end{eqnarray} 
where the $J_n(x)$ are Bessel functions of the first kind, and
\begin{equation}
k_{\rm hh,lh} \equiv \frac{1}{\hbar}\sqrt{\frac{2mE}{\gamma_1 \mp 2\gamma_s}} . 
\end{equation} 
When confinement is present, the eigenstates read
\begin{equation}
\Phi^{F_z}_{\pm}(r,\phi) = a_{\pm}^{F_z}\psi^{F_z}_{{\rm hh},\pm}(r,\phi) + b_{\pm}^{F_z}\psi^{F_z}_{{\rm lh},\pm}(r,\phi) ,
\end{equation} 
where the coefficients $a_{\pm}^{F_z}, b_{\pm}^{F_z}$ and the energies $E$ are to be found from the boundary condition $\Phi^{F_z}_{\pm}(R,\phi) = 0$, resulting in the determinant equations
\begin{eqnarray}
0 &=& J_{F_z\mp 3/2}(k_{\rm hh}R) J_{F_z\pm 1/2}(k_{\rm lh}R) \nonumber \\ & & + 3 J_{F_z\pm 1/2}(k_{\rm hh}R) J_{F_z\mp 3/2}(k_{\rm lh}R).
\end{eqnarray}
By solving the above equations, we find that for $\pm$ the lowest eigenenergy corresponds to $F_z=\mp 1/2$, and the second lowest one to $F_{z}=\pm 1/2$. The associated eigenstates $g_{\pm} \equiv \Phi_{\pm}^{\mp 1/2}$ and $e_{\pm} \equiv \Phi_{\pm}^{\pm 1/2}$ for the transverse motion are found by calculating the coefficients $a_{\pm}^{\mp 1/2}$, $b_{\pm}^{\mp 1/2}$, $a_{\pm}^{\pm 1/2}$, and $b_{\pm}^{\pm 1/2}$, respectively, and serve as the basis states in the effective 1D Hamiltonian. Normalization requires
\begin{gather}
\Braket{g_{\pm}|g_{\pm}} = \int_{0}^{R} dr r\int_{0}^{2 \pi}d\phi \left| g_{\pm} \right|^2 = 1 ,
\end{gather}
and analogously for $e_{\pm}$. It turns out that the excited states are purely heavy-hole-like, $b_{\pm}^{\pm 1/2} = 0$, and we choose the complex phases such that all coefficients are real, with $a_{\pm}^{\mp 1/2} < 0$, $b_{\pm}^{\mp 1/2} > 0$, and $a_{\pm}^{\pm 1/2} > 0$.

\section{Bir-Pikus Hamiltonian}
Referring to holes, the Bir-Pikus Hamiltonian reads
\begin{eqnarray}
H_{\rm BP} &=& -\left( a + \frac{5}{4} b \right) \sum \limits_{i} \epsilon_{ii} + b \sum \limits_{i} \epsilon_{ii} J_i^2 \nonumber \\ 
& & + \frac{2 d}{\sqrt{3}}\Big( \epsilon_{xy} \left\{ J_x, J_y \right\} + \mbox{c.p.}\Big),
\label{BirPikusOfficial}
\end{eqnarray}
where $a$, $b$, and $d$ are the deformation potentials, $\epsilon_{ij} = \epsilon_{ji}$ are the strain tensor elements, $\{A, B\} \equiv (AB + BA)/2$, and $\mbox{``c.p.''}$ stands for cyclic permutations.\cite{birpikus:book} For Ge, the deformation potentials are $b \simeq -2.5\mbox{ eV}$ and $d \simeq -5.0\mbox{ eV}$,\cite{birpikus:book} so that the spherical approximation $d = \sqrt{3} b$ applies. The hydrostatic deformation potential $a$ accounts for the constant energy shift of the VB in the presence of hydrostatic strain, and therefore does not contribute to $\delta(\gamma)$, i.e., the rescaling of the energy gap $\Delta$.

\section{Quantum dot spectrum}
\label{app:QuantumDots}
When the quantum dot length $L$ is much larger than the core radius $R$, Fig.\ \ref{schemeWireQD}, the spectrum can be well approximated using the effective Hamiltonian for extended states. In the absence of external fields, $F_z$ remains a good quantum number and the Hamiltonian
\begin{equation}
H_{\rm LK}^{\rm eff} = \left(
\begin{array}{cccc}
\frac{\hbar^2k_z^2}{2m_g} & -iCk_z & 0 & 0\\
iCk_z & \frac{\hbar^2k_z^2}{2m_e}+\Delta + \delta(\gamma)  &  0 & 0\\
0 & 0 & \frac{\hbar^2k_z^2}{2m_g}& -iCk_z\\
0 & 0 & iCk_z & \frac{\hbar^2k_z^2}{2m_e}+\Delta + \delta(\gamma)
\end{array}
\right),
\end{equation}
here explicitly written out in the basis $\{g_{+}, e_{-}, g_{-}, e_{+}\}$ for illustration purposes, is $2\times 2$ block diagonal with degenerate eigenstates. The subspace $\{g_{+}, e_{-}\}$ corresponds to $F_z = -1/2$, while $\{g_{-}, e_{+}\}$ corresponds to $F_z = +1/2$. Aiming at the quantum dot spectrum, we introduce two complex functions $g_n (z)$ and $e_n (z)$, for which we require
\begin{equation}
\left( \begin{array}{cc}
\frac{\hbar^2k_z^2}{2m_g} & -iCk_z \\
iCk_z & \frac{\hbar^2k_z^2}{2m_e}+\Delta + \delta(\gamma) 
\end{array} \right)
\left( \begin{array}{c}
g_n (z) \\
e_n (z) 
\end{array} \right) = E_n \left( \begin{array}{c}
g_n (z) \\
e_n (z) 
\end{array}\right).
\end{equation}
The associated set of coupled differential equations reads
\begin{eqnarray}
0 &=& -\frac{\hbar^2}{2 m_g}  g_n^{\prime \prime}(z) - C e_n^{\prime}(z) - E_n g_n(z) ,\\
0 &=& - \frac{\hbar^2}{2 m_e}  e_n^{\prime \prime}(z) + C g_n^{\prime}(z) + \left[\Delta + \delta(\gamma) - E_n \right] e_n(z) ,
\end{eqnarray}
and in addition we demand $g_n(0) = e_n(0) = g_n(L) = e_n(L)=0$ due to hard wall confinement at $z=0$ and $z=L$. When the differential equations have been solved, these boundary conditions finally lead to a determinant equation for the eigenenergies $E_n$, which can be analyzed numerically. The results are plotted in Fig.\ \ref{QDsplitting}.

\section{Spin-orbit energy in InAs nanowires}
\label{app:SOenergyInAsNanowires}
For electrons in an electric field $E_x$ along $x$, the Hamiltonian for Rashba SOI is of the form
\begin{equation}
H_{\rm SO}^{\rm el} = \alpha E_x (k_z \sigma_y - k_y \sigma_z),
\end{equation} 
where $\alpha$ is the Rashba coefficient in the conduction band ($\Gamma_6^c$) and $\sigma_i$ are the Pauli matrices for spin 1/2.\cite{winkler:book} In the following, we use the notation $\alpha_x \equiv \alpha E_x$ for illustration purposes. Assuming a nanowire in which the electron moves freely along the $z$ direction with effective mass $m^*$, the Hamiltonian of the system becomes
\begin{equation}
H^{\rm el} = \frac{\hbar^2 k_z^2}{2 m^*} + \alpha_x k_z \sigma_y ,
\end{equation} 
with eigenspectrum
\begin{eqnarray}
E_{\pm} &=& \frac{\hbar^2}{2 m^*}\left( k_z \pm \frac{m^* \left|\alpha_x\right|}{\hbar^2}\right)^2 - \frac{m^* \alpha_x^2}{2 \hbar^2} \nonumber \\
&=& \frac{\hbar^2}{2 m^*}\left( k_z \pm l_{\rm SO}^{-1} \right)^2 - E_{\rm SO}.
\end{eqnarray} 
The spin-orbit length is defined as $l_{\rm SO} \equiv \hbar^2/\left(m^* \left| \alpha_x \right| \right)$, and the SO energy, the energy difference between the band minima and the degeneracy at $k_z =0$, is $E_{\rm SO} = m^* \alpha_x^2/(2 \hbar^2)$, so that
\begin{equation}
E_{\rm SO} = \frac{\hbar^2}{2 m^*} l_{\rm SO}^{-2} . \label{relationESO}
\end{equation}
We can use Eq.\ (\ref{relationESO}) to calculate the spin-orbit energy for InAs wires, where $l_{\rm SO}$ has recently been measured.\cite{fasth:prl07, dhara:09} Using $l_{\rm SO} \simeq 127\mbox{ nm}$ and $m^* \simeq m^*_{\rm bulk} = 0.023\mbox{ }m$,\cite{fasth:prl07} the SO energy in InAs is $E_{\rm SO} \simeq 100\mbox{ $\mu$eV}$. Further experiments confirmed that $l_{\rm SO}$ typically varies between 100 and 200 nm in InAs nanowires,\cite{dhara:09} and in the latter case $E_{\rm SO} \simeq 40\mbox{ $\mu$eV}$ only.

\section{Standard Rashba SOI and Rashba coefficient}
\label{app:RSOIandCoefficient}
Both Ge and Si are inversion symmetric, and thus coupling of Dresselhaus type is absent. However, this does not exclude the conventional Rashba term (RSOI), Eq.\ (\ref{rashbaTermMainText}). Here we briefly outline its derivation; details are described in \mbox{Ref.\ \onlinecite{winkler:book}}. As in Sec.\ \ref{sec:DRSOI}, we assume a constant electric field $E_x$ along the $x$ axis, which, referring to holes, results in the dipole term $H_{\rm ed} = - e E_x x$ as a perturbation added to the potential energy. Accordingly, $H_{\rm ed}$ is added to the multiband Hamiltonian (envelope function approximation), where it appears only on the diagonal, while off-diagonal parts provide the $\bm{k}\cdot\bm{p}$ coupling. Finally, a Schrieffer-Wolff transformation of the multiband Hamiltonian, with focus on the valence band $\Gamma_8^v$, yields the Rashba term
\begin{gather}
H_{\rm SO} = \alpha E_x(k_yJ_z-k_zJ_y), \label{rashbaTermAppendix} \\
\alpha \simeq - \frac{e P^2}{3 E_0^2}, \label{alphaFormula}
\end{gather} 
in third order of perturbation theory, where $\alpha$ is the Rashba coefficient and additional, negligible terms have been omitted. In Eq.\ (\ref{alphaFormula}), $E_0$ is the band gap (direct, $k=0$) between conduction ($\Gamma_6^c$) and valence ($\Gamma_8^v$) band, and $P$ is the corresponding momentum matrix element between the $s$-like $\Gamma_6^c$ and the $p$-like $\Gamma_8^v, \Gamma_7^v$ states.\cite{winkler:book} For Ge, explicit values are $E_0 = 0.90\mbox{ eV}$ and $P = 9.7\mbox{ eV\AA}$,\cite{richard:04} which yields $\alpha \approx -0.4\mbox{ nm}^2 e$.

We can project Eq.\ (\ref{rashbaTermAppendix}) onto the low-energy subspace $\{g_{+}, g_{-}, e_{+}, e_{-}\}$ by calculating the 16 matrix elements. The effective Hamiltonian for RSOI takes on the form
\begin{equation}
H_{\rm R} = H_{\rm SO}^{\rm eff} = \alpha E_x S \bm{\tau}_x \bm{\sigma}_z + \alpha E_x k_z \cdots , 
\label{eff_ham_x_SOI}
\end{equation}
where $S = \bra{g_{+}}k_y J_z\ket{e_{+}}$.
This Hamiltonian has two effects: first, it features a constant coupling between the $g$ and $e$ states, and second, it provides a term which is linear in $k_z$ and mixes the spin blocks. The latter is absent at $k_z = 0$, so that only the constant term $\alpha E_x S \bm{\tau}_x \bm{\sigma}_z$ contributes for small $k_z$; this is of the same form as the direct Rashba SOI $H_{\rm DR} = e E_x U \bm{\tau}_x \bm{\sigma}_z$ (DRSOI) resulting from dipolar coupling. Finally, we note that 
\begin{equation}
\frac{e E_x U}{\alpha E_x S} \simeq -1.1 \frac{R^2}{\mbox{nm}^2} 
\end{equation}
for Ge, so that the DRSOI dominates RSOI by one to two orders of magnitude in typical Ge/Si nanowires of $\mbox{5-10 nm}$ core radius.

\section{Coupling to magnetic fields}
\label{app:MagneticField}
In Eqs.\ (\ref{effBalongZ}) and (\ref{effBalongX}), we show the effect of external magnetic fields on the low-energy sector for fields applied along ($z$) and perpendicular ($x$) to the nanowire, respectively. Below, the explicit values for $Z_i$ and $X_i$ are listed,
\begin{equation}
\begin{array}{rclrcl}
Z_1 &=& 0.75, & \hspace{0.6cm} X_1 & = & 2.72, \\  
Z_2 &=& -0.81, & \hspace{0.6cm} X_2 & = & 0.17, \\
Z_3 &=& 2.38\mbox{ }R, & \hspace{0.6cm} X_3 & = & 8.04\mbox{ }R, 
\end{array} \label{bFieldZiAndXi}
\end{equation} 
using the parameters $\gamma_1 = 13.35$, $\gamma_s = 5.11$, and $\kappa = 3.41$ for Ge.\cite{lawaetz:prb71} 



\begin{thebibliography}{}
\bibitem{commentLK}{Equation (\ref{LK_spherical}) is valid for $[k_i, k_j] = 0$ only. Magnetic field effects are included via the form derived in Ref.\ \onlinecite{luttinger:pr56}.}
\bibitem{commentPhases}{Phases in the off-diagonals depend on the actual representation of the spin operators and eigenstates. The details are summarized in Appendixes \ref{app:Representation} and \ref{app:BasisStates1D}.}
\end{thebibliography}
\end{document}